\documentclass[aps,prl,floatfix,epsfig,twocolumn,showpacs,preprintnumbers]{revtex4}
\usepackage{amssymb}
\usepackage{graphicx}
\usepackage{amsmath}

\input{tcilatex}

\begin{document}

\title{Many-Body Electronic Structure of Americium metal}
\author{Sergej Y. Savrasov$^a$, Kristjan Haule$^{b}$, Gabriel Kotliar$^{b}$}
\affiliation{$^{a}$Department of Physics, New Jersey Institute of Technology, Newark, NJ
07102, USA}
\affiliation{$^{b}$Department of Physics and Center for Material Theory, Rutgers
University, Pscataway, NJ 08854, USA}
\date{\today}

\begin{abstract}
We report computer based simulations of energetics, spectroscopy and
electron-phonon interaction of americium using a novel spectral density
functional method. This approach gives rise to a new concept of a many-body
electronic structure and reveals the unexpected mixed valence regime of Am 5f%
$^{6}$ electrons which under pressure acquire the 5f$^{7}$ valence state.
This explains unique properties of Am and addresses the fundamental issue of
how the localization delocalization edge is approached from the localized
side in a closed shell system.
\end{abstract}

\date{\today}
\maketitle

\input{epsf}

Artificially produced from Plutonium-239 in 1944, and widely used in smoke
detectors Americium is the first transuranic actinide where 5f$^{6}$
electrons become localized and form a closed relativistic subshell. Its
recent high-pressure studies\cite{1} have drawn much attention as
understanding volume behavior in actinides systems has important
consequences on their storage and disposal. They have revealed that Am
undergoes a series of structural phase transitions (denoted hereafter as I,
II, III, and IV) and reproduces at least two of the structures of another
mysterious element, Plutonium, which links the physical behavior of all
actinides materials to our fundamental understanding of bonding between
their 5f-electrons. At ambient pressure Am I behaves as an ordinary metal
with slightly enhanced electrical resistivity $\rho $(T=300K)=68 $\mu \Omega
\times cm$ and no sign of ordered or disordered magnetism. This is
standardly understood as a manifestation of $_{7}$F$^{0}$ ground state
singlet of 5f$^{6}$ atomic configuration. However, the resistivity of Am
raises almost an order of magnitude and reaches its value of 500 $\mu \Omega
\times cm$ at the orthorhombic structure of Am IV which is realized at
pressures P above 16 GPa. The most prominent feature of the pressure $P$ vs.
volume $V$ behavior is the existence of two distinct phases: the
\textquotedblleft soft\textquotedblright\ one which occurs in Am I through
III as well as another \textquotedblleft hard\textquotedblright\ phase
realized in Am IV. On top of that a superconductivity in Am was first
predicted\cite{2} and then discovered\cite{3} with Tc raising from 0.5K in
Am I to 2.2K in Am II, falling slightly in Am III and then exhibiting a
sharp maximum in phase IV\cite{4}.

Understanding this unique behavior is a fundamental challenge in searching
for a unified theory of actinides as the pressure driven delocalization of
electrons is approached here from the localized side which is very different
from Pu where originally delocalized electrons become localized with
increasing volume. Thus, simple model Hamiltonians which contain qualitative
features to produce complex energy landscapes with multiple solutions in
open shell systems cannot be employed for studies of closed shell materials:
without incorporating realistic structures in the calculation, there is no
hint of bistability in the model Hamiltonian approach.

To address these issues in this work we introduce a novel many-body
electronic structure method which allows us to uncover the physics of Am. It
is based on dynamical mean field theory (DMFT), a modern many body technique
for treating strongly correlated electronic systems in a non-perturbative
manner\cite{5,6} and at the same time has computational efficiency
comparable with ordinary electronic structure calculations thus allowing us
to deal with complicated crystal structures of real solids by
self-consistent many-body calculations. Our new method considers the local
Green function $G_{loc}(\omega )$ as a variable in the total energy
functional and can be viewed as spectral density functional theory\cite%
{7,8,9}. The advantage of such formulation as compared to original density
functional theory\cite{10} is a simultaneous access to energetics and local
excitation spectra of materials with arbitrary strength of the local Coulomb
interaction U.

DMFT based spectral density functional approach requires self-consistent
solutions of the Dyson equations

\begin{equation}
\lbrack \omega -H_{0}(\mathbf{k})-\Sigma (\omega )]G(\mathbf{k},\omega )=1
\label{Eq1}
\end{equation}
for the one-electron Green function $G(\mathbf{k},\omega )$. The poles of
its momentum integrated $G_{loc}(\omega )$ contain information of the true
local spectra of excitations. Here $H_{0}(\mathbf{k})$ is the effective
one-electron Hamiltonian while $\Sigma (\omega )$ is a local self-energy
operator whose energy dependence makes the solution computationally very
expensive. This so far has restricted applications of this promising
many-body approach either to non-self-consistent determinations of spectra%
\cite{11} or to materials with simple crystal structures\cite{8,12,13}.

Our new approach greatly improves the speed of the calculation by
recognizing that a signature of strong correlation effect results in
appearance of several distinct features or satellites in the excitation
spectrum. The exact self-energy of an interacting system can always be
represented by a pole expansion of the form

\begin{equation}
\Sigma (\omega )=\Sigma (\infty )+\sum_{i}\frac{W_{i}}{\omega -P_{i}}
\label{Eq2}
\end{equation}%
Remarkably, that such form of the self-energy allows us to replace the
non-linear (over energy) Dyson equation by a linear Schroedinger--like
equation in extended subset of \textquotedblleft pole
states\textquotedblright . This is clear due to a mathematical identity 
\begin{widetext}
\begin{equation}
\left( 
\begin{array}{cc}
\omega -H_{0}(\mathbf{k})-\Sigma (\infty  ) & \sqrt{W} \\ 
\sqrt{W} & \omega -P%
\end{array}%
\right) ^{-1}=\left( 
\begin{array}{cc}
[\omega -H_{0}(\mathbf{k})-\Sigma (\infty )-\sqrt{W}(\omega -P)^{-1}\sqrt{W}]^{-1}
& \ldots  \\ 
\vdots  & \ddots 
\end{array}%
\right)   \label{Eq3}
\end{equation}%
\end{widetext} which relates our original matrix inversion required to find $%
G(\mathbf{k},\omega )$ (first element in the matrix from the right) to the
matrix inversion in the extended \textquotedblleft pole
space\textquotedblright .

The key insight is that the above form of the self energy with a few poles
captures\cite{9,14} all the central features of a correlated system and is
an excellent approximation to the Greens function of the system for the
purposes of obtaining the total energy. This has an important implication
for the calculation of the electronic structure of the strongly correlated
material: once pole expansion of the self-energy is established, the
spectral density functional theory reduces to solving a \textquotedblleft
Kohn Sham--like\textquotedblleft\ system of equations in an augmented space.
The eigenstates here describe major atomic multiplet transitions as well as
delocalized parts of the electronic states by separate auxiliary wave
functions. Each wave function is not normalized to unity since it describes
only part of the spectral weight for the electron leaving in the vicinity of
a given energy, however the integral spectral weight over all energies is
correctly normalized to one. While a self energy with a small number of
poles may not capture subtle physics of damping important for incoherent
excitations, the method gives us directly the dispersions of these spectral
features, which are measurable in angle resolved photoemission.

Thus, the concept of the electronic structure is generalized to a strongly
correlated situation. It ideally suits the description of such subtle regime
as the proximity to the Mott transition where atomic multiplet structure
appears simultaneously with strongly renormalized quasiparticle bands, a
regime where traditional electronic structure methods fail.

Here, we study the properties of Am under pressure using this newly
implemented matrix expansion algorithm for spectral density functional
calculations within a full potential version of the linear muffin-tin
orbital method \cite{15}. In this approach, the s,p,d electrons are assumed
to be weakly correlated and well described within such popular
approximations to the density functional theory as the local density
approximation (LDA) including gradient corrections (GGA). The correlated f
electrons require dynamical treatment using DMFT. Both, the spin-orbit as
well as the Hund's couplings are competing in Americium and need to be taken
into account. The former is a one-body term and enters through the LDA
Hamiltonian, while the second is contained in the local Coulomb repulsion,
which is conveniently expressed via Slater constants F$^{(i)}$. The value of
the most important term F$^{(0)}$=U is around 4.5 eV which is suggested from
various atomic spectroscopy data and our previous studies of Plutonium. For
the remaining constants we take the atomic values F$^{(2)}$=7.2 eV, F$^{(4)}$%
=4.8 eV, F$^{(6)}$=3.6 eV\cite{16}.

For the purpose of the total-energy calculation, the f-electron self-energy
is approximated by its atomic value which is obtained by the exact
diagonalization technique. This is known as the Hubbard I approximation. The
probabilities to find the f-shell in its given many body state are directly
accessed within this method and give us the insight into valence of the
material.

Our calculation reproduces the well known fact that the f electrons in Am at
zero pressure exists in a f$^{6}$ $_{7}$F$^{0}$ configuration. This is
illustrated in Fig. 1 by plotting density of states and energy bands
reflecting the atomic multiplet transitions which demonstrate our novel
matrix expansion algorithm. Our calculated one-electron density of states
shown on the right consists of several distinct features related to f$%
^{6}\rightarrow $f$^{5}$ electron removal and f$^{6}\rightarrow $f$^{7}$
electron addition processes. One can see that occupied part of the spectrum
is well compared with the available photoemission experiments\cite{17}, thus
advancing previous bulk-surface interpretation\cite{18} as well as density
functional based calculations\cite{19,20}. The idea of our new method is to
model three major satellites related to $_{7}$F$^{0}\rightarrow $ $_{6}$H$%
^{5/2}$, $_{7}$F$^{0}\rightarrow $ $_{8}$S$^{7/2}$ and $_{7}$F$%
^{0}\rightarrow $ $_{6}$P$^{7/2}$ transitions with two pole self-energy and
resolve them as many-body energy bands. This is illustrated on left part of
Fig. 1 where the method is seen to capture the spectral weight related to
electron removal and addition processes by a set of eigenstates located near
--3 eV binding energy (blue) and by two sets of eigenstates located at +1
(green) and +3 eV (red). A simplified fcc structure with equilibrium atomic
volume of Am I was used to generate the data in Fig 1.

\begin{figure}[tbp]
\includegraphics*[height=3.0in]{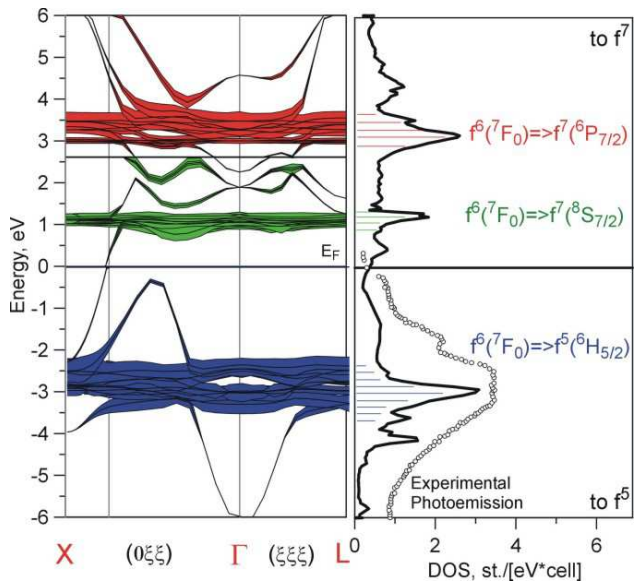}
\caption{Calculated many body set of energy bands (left) which models the
one-electron density of states (right) of Americium metal. Experimental
photoemission spectrum\protect\cite{17} is also shown by open circles.}
\label{Fig1}
\end{figure}

The computational speed gained by this algorithm allows us to study
complicated crystal structures of Am. In particular, the existence of soft
and hard phases in its equation of state can be predicted via our
self-consistent total energy calculations. This is illustrated in Fig. 2
where P(V) behavior reconstructed from the total energy data of phases I
through IV is plotted and compared with the recent experiment. For Am I we
predict the equilibrium volume equal to 27.4 \AA $^{3}$/atom which is only
7\% less than the experiment together with the bulk modulus equal to 450
kBar close to the experimentally deduced values lying within 400-450 kBar.
The pressure ranges of all other structures are correctly reproduced. A
compressibility of highly pressurized Am IV structure is found to behave
similarly to experimentally observed \textquotedblleft
hard\textquotedblright\ phase which indicates that f-electrons start
participating in bonding. Some discrepancy between the calculated and the
measured compressibility of Am III can be observed in Fig.2. This is likely
to be due to simplified impurity solver or due to the uncertainties in the
estimates of the Hubbard U.

To gain theoretical insight and understand the origin of
localization-delocalization transition we now discuss the behavior of the
electronic structure under pressure. To see how the increase of
hybridization among f-electrons affects the physical properties of Am, we
carry out subsequent refined calculations, by replacing the Hubbard I
approximation by a more precise one-crossing approximation (OCA) method\cite%
{13,21} to solve the Anderson impurity problem. Due to numerical complexity
of the approach, we calculate only spectral functions using the fcc
structure of Am and omit self-consistent determination of the energy.

\begin{figure}[tbp]
\includegraphics*[height=2.8in]{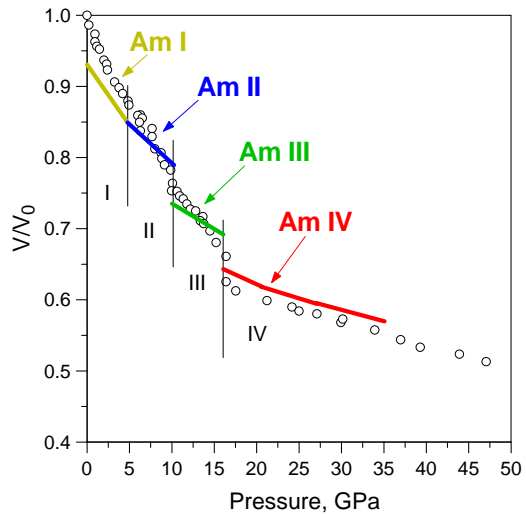}
\caption{Calculated equation of state for various crystal structures of Am
metal. Experimental data1 are shown by solid cicrles.}
\label{Fig2}
\end{figure}

Fig. 3 shows the density of states for Am at three different volumes V=V$%
_{0} $,V=0.76V$_{0}$ and V=0.63V$_{0}$ which covers the lattice spacing of
the entire phase diagram discussed above. Upon compression, the remarkable
effect is observed as peak near the Fermi level gets pushed down while a
resonance (small shoulder) starts forming at E$_{f}$ and becomes more
pronounced with increasing pressure. The f$^{6}$ ground state of the atom
starts admixing an f$^{7}$ configuration with a very large total spin of
J=7/2. Due to hybridization with the spd bands, this large spin gets
screened thus lowering the energy of the system. This is the famous Kondo
mechanism, and the energy gain increases as the hybridization increases by
applying pressure.

The admixture of the f$^{7}$ configuration is counterintuitive. Naively one
expects that application of pressure results in lowering the Fermi level in
the spd band (which contains only 3 electrons) which then moves towards the
f level. This reduces the occupancy of the f, admixing an f$^{5}$
configuration, an effect that is known to induce mixed valence in Sm
compounds \cite{211}. Our first principles calculations reveal that while
the position of the bare f level in Am indeed moves upwards relative to the
Fermi level, the energy to absorb an electron to reach f$^{7}$ configuration
is much smaller than the energy to remove an f electron and transfer it to
the Fermi level hence reaching the f$^{5}$ configuration. Application of
pressure reduces the energy of the f$^{7}$ configuration in the bath of spd
electrons by the gain in hybridization, and the resulting energy gain is
sufficient to compensate the increase in the distance between the bare f
level and the Fermi level. In this way, the valence is also changed from 3
to approximately 2.8 under extreme pressure. This confirms the attribution
of the rise in resistivity to mixed valence and provides the microscopic
mechanism of this phenomenon. At normal pressure, the f shell is essentially
closed and unable to scatter the conducting bands. This results in small
resistivity. Under pressure, the reaction f$^{6}$ + spd $\rightarrow $ f$%
^{7} $ becomes more energetically allowed resulting in the growth of
resistivity.

\begin{figure}[tbp]
\includegraphics*[height=2.8in]{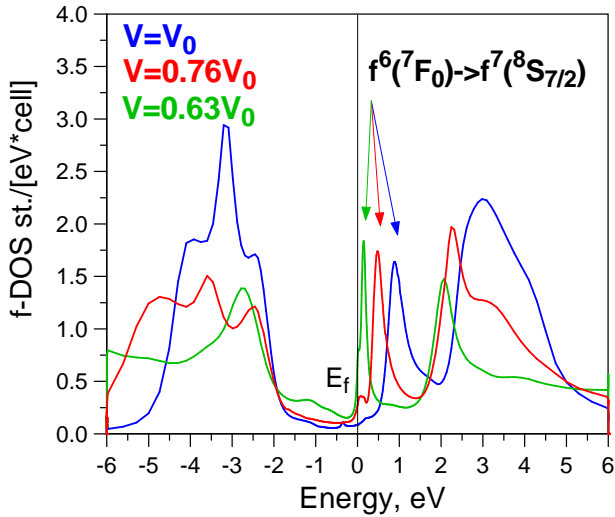}
\caption{Pressure dependence of Am density of states calculated using the
one crossing approximation method.}
\label{Fig3}
\end{figure}

It is interesting to compare the DMFT predictions for Am with the
predictions of the density functional theory. The non-magnetic GGA
calculation falls catastrophically in reproducing the theoretical
equilibrium volume of the soft phase and underestimates it by about 50\%.
When spin polarization is allowed, the GGA eventually recovers most of this
error but converges to the wrong magnetic state with its total (spin plus
orbital) moment of about 6 Bohr magneton \cite{19,20}. This prediction is at
odds with the experimentally established f$^{6}$ ground state singlet $_{7}$F%
$^{0}$. Similar findings have been reported when applying a disordered local
moment method \cite{201}. This error is the result of neglecting an
important correlation effect. The Kohn Sham spectrum of Am describes an f
level with a small spin orbit splitting between f$^{5/2}$ and f$^{7/2}$ (of
the order of 1 eV) which leads to two energy bands located just near E$_{f}$
and which are unstable against magnetism. In Am the Coulomb interaction
increases this splitting by the value of U (of the order of 4 eV) and leads
to the atomic-like $_{7}$F$^{0}$ many-body state. This stabilization of the
closed shell due to the Hubbard interaction U is absent in the band
calculation.

We finally estimate the superconducting critical temperature by computing
from first principles \cite{22} the electron-phonon coupling of the
electrons in the presence of correlations. For this purpose we have extended
a newly developed dynamical mean field based linear response method, which
has previously proven to provide accurate phonon spectra in correlated
systems \cite{23,24}. We estimate the coupling constant which comes out to
be sufficiently high ($\sim $0.5) to predict superconductivity of the order
of 1 K. The occurrence of the first maximum in experimental T$_{c}$ vs
pressure dependence, can then be understood as the result of the variation
of the spd density of states which first increases as a result of a band
structure effect but then eventually decreases as the hybridization with the
f electron grows with the increase of mixed valence.

To summarize, here we provided a first order picture of electronic
properties of Am metal but the necessity of its further studies is apparent.
These require extensions of our methods to evaluate the electron--phonon and
the Coulomb interactions among quasiparticles in a full fledged mixed
valence state and will be carried out in the future work.

Support by the DOE grant DE-FG02 99ER45761, the NSF grants 0238188, 0312478,
0342290 and US DOE Computational Material Science Network is gratefully
acknowledged.


\begin{thebibliography}{99}
\bibitem{1} S. Heathman, R. G. Haire, T. Le Bihan, A. Lindbaum, K. Litfin,
Y. M\'{e}resse, and H. Libotte, Phys. Rev. Lett. 85, 2961 (2000).

\bibitem{2} B. Johansson and A. Rosengren, Phys. Rev. B \textbf{11}, 2836
(1975).

\bibitem{3} J.L. Smith and R.G. Haire, Science \textbf{200}, 535(1978).

\bibitem{4} J.-C. Griveau, J. Rebizant, and G. H. Lander, G. Kotliar, Phys.
Rev. Lett. \textbf{94}, 097002 (2005).

\bibitem{5} Gabriel Kotliar and Dieter Vollhardt, Physics Today \textbf{57},
53 (2004).

\bibitem{6} A.Georges, G. Kotliar, W. Krauth and M. J. Rozenberg, Rev. Mod.
Phys. \textbf{68}, 13 (1996).

\bibitem{7} R. Chitra, and G. Kotliar, Phys. Rev. B \textbf{63}, 115110
(2001).

\bibitem{8} S. Savrasov, G. Kotliar, and E. Abrahams, Nature \textbf{410},
793 (2001).

\bibitem{9} S. Y. Savrasov, G. Kotliar, Phys. Rev. B \textbf{69}, 245101
(2004).

\bibitem{10} For a review, see, e.g., Theory of the Inhomogeneous Electron
Gas, edited by S. Lundqvist and S. H. March (Plenum, New York, 1983).

\bibitem{11} For recent applications, see, e.g., S. Biermann, A. Poteryaev,
A. I. Lichtenstein, and A. Georges, Phys. Rev. Lett. \textbf{94}, 026404
(2005).

\bibitem{12} K. Held, G. Keller, V. Eyert, D. Vollhardt, and V. I. Anisimov,
Phys. Rev. Lett. \textbf{86}, 5345 (2001).

\bibitem{13} K. Haule, V. Oudovenko, S. Y. Savrasov, G. Kotliar, Phys. Rev.
Lett. \textbf{94}, 036401 (2005).

\bibitem{14} S. Y. Savrasov, V. Oudovenko, K. Haule, D. Villani, G. Kotliar,
Phys. Rev. B \textbf{71}, 115117 (2005).

\bibitem{15} S. Y. Savrasov, Phys. Rev. B \textbf{54}, 16470 (1996).

\bibitem{16} W. T. Carnall and B. G. Wybourne, J. Chem. Phys. \textbf{40},
3428 (1964).

\bibitem{17} J. R. Naegele, L. Manes, J. C. Spirlet, and W. Muller, Phys.
Rev. Lett. \textbf{52}, 1834 (1984).

\bibitem{18} N. Martensson, B. Johansson, J. R. Naegele, Phys. Rev. B\textbf{%
\ 35}, 1437 (1987).

\bibitem{19} P. Soderlind, R. Ahuja, O. Eriksson, and B. Johansson, J. M.
Wills, Phys. Rev. B \textbf{61}, 8119 (2000).

\bibitem{20} M Penicaud, J. Phys.: Condens. Matter \textbf{17}, 257 (2005).

\bibitem{201} A. M. N. Niklasson,J. M. Wills, M. I. Katsnelson, I. A.
Abrikosov, O. Eriksson, and Borje Johansson, Phys. Rev. B \textbf{67},
235105 (2003).

\bibitem{21} K. Haule, S. Kirchner, J. Kroha, and P. W\"{o}lfle, Phys. Rev.
B \textbf{64}, 155111 (2001).

\bibitem{211} A. Svane, V. Kanchana, G. Vaitheeswaran, G. Santi, W. M.
Temmerman, Z. Szotek, P. Strange, and L. Petit, Phys. Rev. B \textbf{71},
045119 (2005).

\bibitem{22} S. Y. Savrasov, D. Y. Savrasov and O. K. Andersen, Phys. Rev.
Lett.\textbf{\ 72}, 372 (1994).

\bibitem{23} X. Dai, S. Y. Savrasov, G. Kotliar, A. Migliori, H. Ledbetter,
E. Abrahams, Science \textbf{300}, 953 (2003).

\bibitem{24} S. Y. Savrasov, G. Kotliar, Phys. Rev. Lett. \textbf{90},
056401 (2003).
\end{thebibliography}
\end{document}